\documentclass[aps,prb,preprint,groupedaddress]{revtex4-1}
\usepackage{graphicx}
\usepackage{bm}
\usepackage{times}
\usepackage{colordvi,epsf}

\begin{document}

\title{How to reveal metastable skyrmionic spin structures
by spin-polarized scanning tunneling microscopy}
\author{B.~Dup\'e, C.~N.~Kruse, T.~Dornheim and S.~Heinze}
\address{Institute of Theoretical Physics and Astrophysics, Christian-Albrechts-Universit\"at zu Kiel,
24098 Kiel, Germany}

\date{\today}

\begin{abstract}
We predict the occurrence of metastable skyrmionic spin structures such as antiskyrmions and higher-order skyrmions in ultra-thin transition-metal films 
at surfaces using Monte Carlo simulations based on a spin Hamiltonian parametrized from density functional theory calculations.
We show that such spin structures will appear with a similar contrast in spin-polarized scanning tunneling microscopy (SP-STM) images. 
Both skyrmions and antiskyrmions display a circular shape for out-of-plane magnetized tips and a two-lobe butterfly contrast for in-plane tips.
An unambiguous distinction can be achieved by rotating the tip magnetization direction without requiring the information of all components of the magnetization.  
\end{abstract}

\maketitle

\section{Introduction}

Magnetic skyrmions were recently observed via neutron diffraction in bulk chiral magnets such as MnSi~\cite{muhlbauer2009skyrmion} and in the multiferroic material
Cu$_2$OSeO$_3$~\cite{Seki1304201}. Currently, they are attracting an enormous attention due to their 
stability~\cite{Bogdanov1994,Bogdanov1994a} and their displacement speed upon applying electrical currents
which makes them suitable for technological applications~\cite{nnano.2013.29,Nagaosa:13.1}. Real space observation of skyrmions in FeGe and Fe$_{0.5}$Co$_{0.5}$Si thin films has become possible using Lorentz microscopy and magnetic force microscopy~\cite{yu2010real,Yu2010_FeGe,PhysRevLett.108.267201,Milde2013} and 
more recently in transition-metal films using spin-polarized low-energy electron microscopy~\cite{Chen2015} and
magneto-optical Kerr effect (MOKE) measurements~\cite{Jiang2015}. 
In ultra-thin films of a few monolayers, the skyrmion diameter can shrink down to a few nanometers and spin-polarized scanning tunneling microscopy (SP-STM) 
\cite{Bode2003,RevModPhys.81.1495} is a powerful tool for their observation and manipulation~\cite{nphys2045,Romming-2013aa,Romming2015}.

SP-STM is sensitive to the projection of the local magnetization density of states of the sample onto the magnetization direction of the tip
\cite{PhysRevLett.86.4132} and does not allow a direct determination of the three magnetization components in a single measurement. 
However, in most experimental setups it is not possible to continuously rotate the tip
magnetization direction and conclusions have to be drawn from SP-STM experiments performed with only one or two tip magnetization directions.
Such measurements only allow a partial determination of the spin structure~\cite{Romming-2013aa,Bergmann:2014aa,Phark2014}. It is therefore essential to know (i) if the skyrmion ground states and some metastable states can be differentiated via simple SP-STM experiments (based on one or two tip magnetization directions) and (ii) to establish a 
clear proposal in order to discriminate between the different possible chiral spin structures via SP-STM.

\section{Methods}
The occurrence of metastable skyrmionic spin structures is studied in a single atomic layer of Pd in fcc stacking on the fcc monolayer Fe on the Ir(111) 
surface denoted as Pd(fcc)$/$Fe$/$Ir(111). This system has been studied experimentally using SP-STM \cite{Romming-2013aa,Romming2015} and from first-principles 
calculations \cite{Dupe:2014aa,Simon2014} which allow to understand the transition from a spin spiral to a skyrmion and a ferromagnetic (FM) phase in an external 
magnetic field. We numerically solve the spin Hamiltonian using Monte-Carlo (MC) simulations
with parameters obtained from density functional theory calculations~\citep{Dupe:2014aa}:

\begin{eqnarray}
H & = & -\sum_{ij} J_{ij} \mathbf{M}_i \cdot \mathbf{M}_j -  \sum_{ij} \mathbf{D}_{ij} \cdot \left( \mathbf{M}_i \times \mathbf{M}_j \right) \nonumber \\
{} & + & \sum_{i} K (M_i^z)^2 + \sum_{i} \mathbf{B} \cdot \mathbf{M}_i
\label{eqn:H} 
\end{eqnarray} 

with exchange constants $J_{ij}$, the vector ${\mathbf D}_{ij}$ of the Dzyaloshinskii-Moriya (DM) interaction, 
the magnetocrystalline anisotropy $K$ and an external magnetic field~\cite{coeff}.

We have obtained metastable states in MC simulations by relaxing a super cell of 100$\times$100 spins on a two-dimensional hexagonal lattice 
starting from a random spin configuration at 1~K under a magnetic field of 20~T with a standard Metropolis algorithm. 
At this field value we are in the region where the skyrmions are metastable in the ferromagnetic background~\cite{Dupe:2014aa}. 

\begin{figure}[htp]
\centering
\includegraphics[width=\columnwidth]{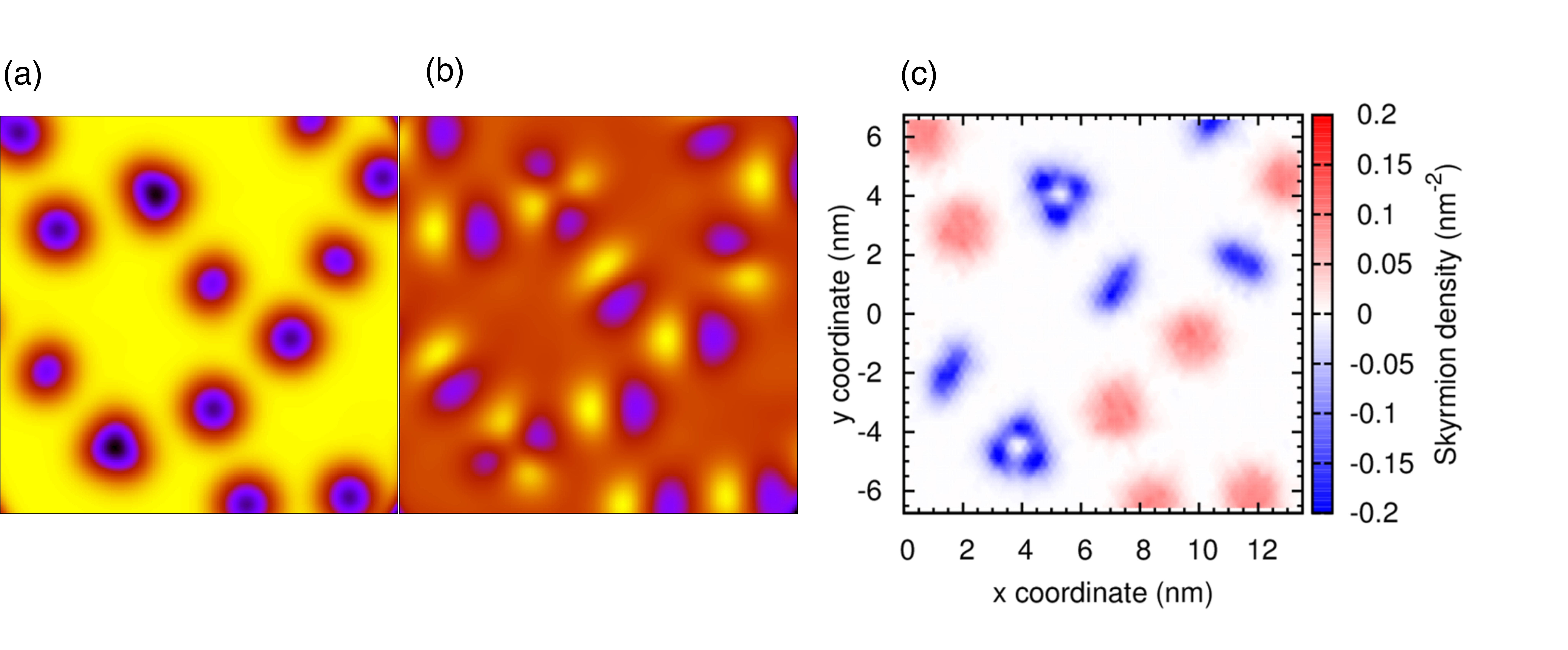}
\caption{Simulated SP-STM image with an out-of-plane (a) and an in-plane (b) magnetized tip at $z=8$~\AA~from the surface for a spin structures of~$13\times13$~nm$^2$ obtained from MC simulations for Pd/Fe/Ir(111). (c) Skyrmion density of the spin structure (color contrast). The spin structure was obtained with a super cell of 100$\times$100 spins on a hexagonal lattice at a temperature of
1~K after a relaxation with $10^7$ MC relaxation starting from a random spin configuration at a magnetic field of $B=20$~T, i.e.~in the region of the phase diagram
in which the ferromagnetic state is the ground state.}
\label{fig:relax}
\end{figure}

We have simulated SP-STM images of the spin structures obtained from MC using the model described in Ref.~\cite{Heinze2006}. The tunneling current is given by
\begin{equation}
	\label{tunnelcurrent}
	I({\mathbf R}_{\rm T}) \propto \sum\limits_{\alpha}
	\left(1+P_{\rm T} P_{\rm S} \cos \theta_\alpha \right) h({\mathbf R}_{\rm T} - {\mathbf R}_\alpha),
\end{equation}
where ${\mathbf R}_{\rm T}$ is the tip position, the sum extends over all surface
atoms $\alpha$, the vacuum tail of a spherical atomic wave function is
approximated by $h({\mathbf r})=\exp{(-2\kappa |{\mathbf r}|)}$, and the decay
constant is given by $\kappa=\sqrt{2m\phi/\hbar^2}$ with the work function
$\phi$. $P_{\rm S}$ and $P_{\rm T}$ denote the spin-polarization of sample and
tip atoms, respectively, and $\theta_\alpha$ is the angle of
the magnetization of atom $\alpha$ with respect to the tip
magnetization direction ${\mathbf m}_{\rm T}$.

\section{Results}

Figure~\ref{fig:relax}(a) shows a simulated SP-STM image with an out-of-plane magnetized tip ($P_{\rm eff}=P_{\rm T} P_{\rm S}=0.4$) of the spin structure at $z=8$~\AA~from the surface. The image shows a brighter contrast for the FM background with several black spots. All darker spots have a round shape and could correspond to skyrmion spin structures. However, when the tip magnetization is changed from out-of-plane to in-plane (Fig.~\ref{fig:relax}(b)), the simulated SP-STM image shows two types of contrast compatible with recent observation of skyrmion~\cite{Romming2015}. The first contrast has a two-lobe pattern with one brighter and one darker side. The lobes can be aligned along the x axis or are rotated with respect to it. The second type of contrast has four lobes and appears seldom. It does not seem to have a preferred alignment.

The topological character of different spin structures is given by their winding or skyrmion number:
\begin{equation}
 S = \frac{1}{4 \pi} \int \textbf{m} \cdot \left( \frac{\partial \textbf{m} }{\partial x} \times \frac{\partial  \textbf{m} }{\partial y} \right) \, dx \, dy
\label{eqn:S} 
\end{equation}
where $\mathbf{m}$ is the unit vector of the local magnetization and $S$ can take only integer values.
Figure~\ref{fig:relax}(c) shows the skyrmion density as expressed by the integrand of Eq.~(\ref{eqn:S}) as a color contrast. Unfortunately, this quantity is not accessible in experiments but it can differentiate the different spin structures. The two-lobe patterns correspond to a skyrmion (red contrast), which has a positive skyrmion density that integrates to~$S=1$, or to an anti-skyrmion with a negative skyrmion density ($S=-1$) (blue contrast). The four-lobe contrast has a triangular shaped skyrmion density that integrates to the value of~$S=-2$ suggesting the presence of a metastable higher order skyrmion in this system. Experimentally, the different spin structures can be obtained by freezing an ultra-thin film sample from high temperatures to 
a temperature e.g. 1~K under magnetic field or by locally heating the sample ~\cite{Koshibae_NatureComm2014} 
e.g.~with an electrical current while maintaining the sample at a low temperature.

\begin{figure}[thp]
\centering
\includegraphics[width=\columnwidth]{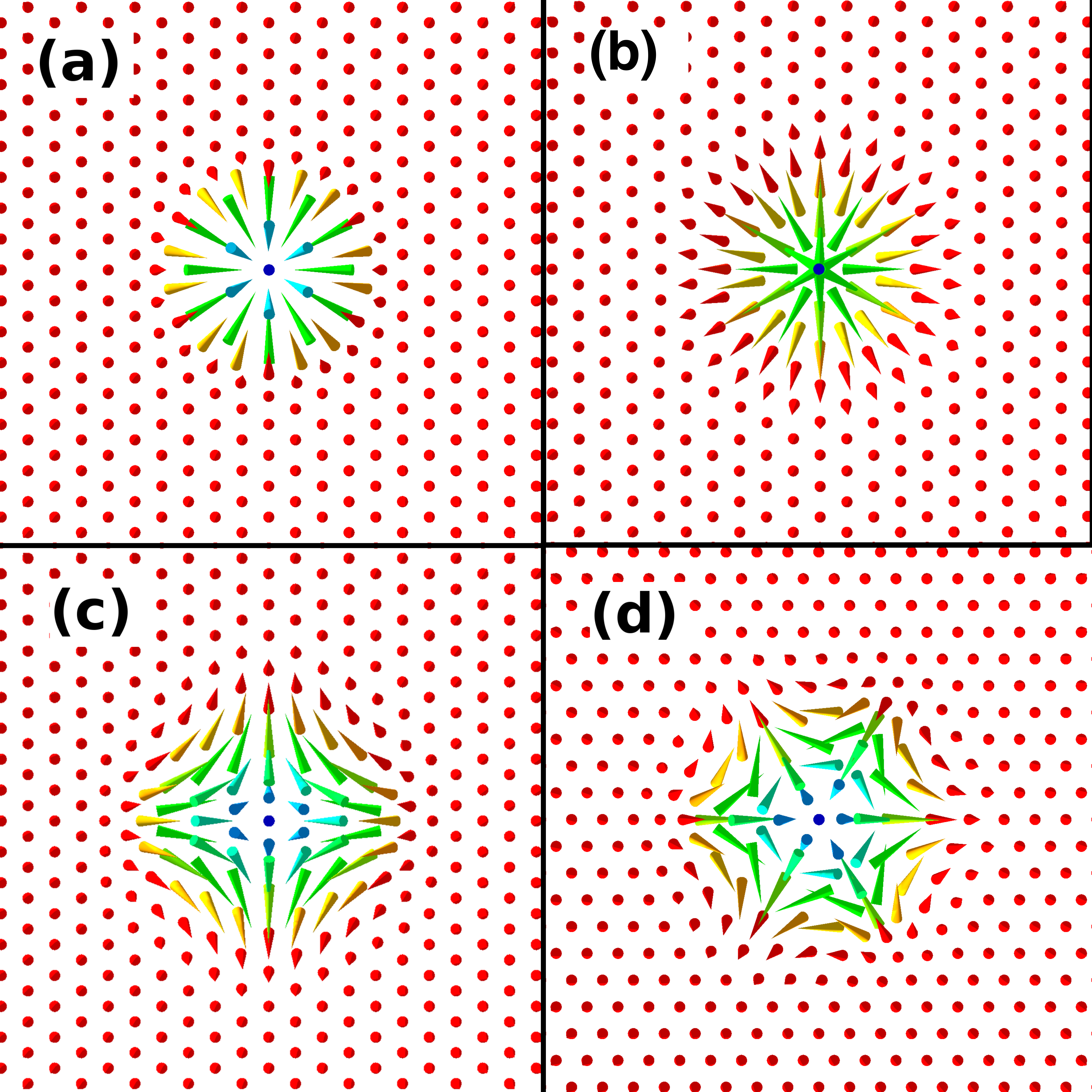}
\caption{Metastable chiral spin textures relaxed at $T=0$~K and $B=20$~T for Pd(fcc)$/$Fe$/$Ir(111). (a) Right-handed skyrmion ($S=1$). (b) Left-handed skyrmion ($S=1$). (c) Antiskyrmion ($S=-1$). (d) Higher-order antiskyrmion ($S=-2$). Depending on the angle $\theta$ of the magnetic moment with respect to the $z$-axis
perpendicular to the film, the color of the arrows changes from red for $\theta=0$ (or $\bf m $ pointing upwards) to blue for $\theta=\pi$ (or $\bf m $ pointing downwards). Green arrows ($\theta=\pi/2$) correspond to in-plane spins.}
\label{fig:comparaison}
\end{figure}

In order to compare the different states in detail, we show in Fig.~\ref{fig:comparaison} the distinct spin structures identified
in Fig.~\ref{fig:relax} in separate panels. Fig.~\ref{fig:comparaison}(a) shows a right-handed skyrmion i.e.~$S=1$
which is metastable for Pd(fcc)$/$Fe$/$Ir(111) at 
magnetic field values higher than 16~T~\cite{Dupe:2014aa}. 
For completeness, we also consider a left-handed skyrmion, Fig.~\ref{fig:comparaison}(b), which has a skyrmion number of $S=1$ as well
but exhibits an opposite chirality and is unstable.
Fig.~\ref{fig:comparaison}(c) shows an antiskyrmion ($S=-1$) that is characterized by a change of chirality for two high symmetry directions, i.e.~the rotational sense changes from right- to left-handed every $90^{\circ}$. Fig.~\ref{fig:comparaison}(d) displays a higher-order antiskyrmion with $S=-2$ which was recently also reported in 
Ref.~\cite{Leonov2015}. In that case, the rotational sense changes every $60^{\circ}$.

\begin{figure}[thp]
\centering
\includegraphics[width=\columnwidth]{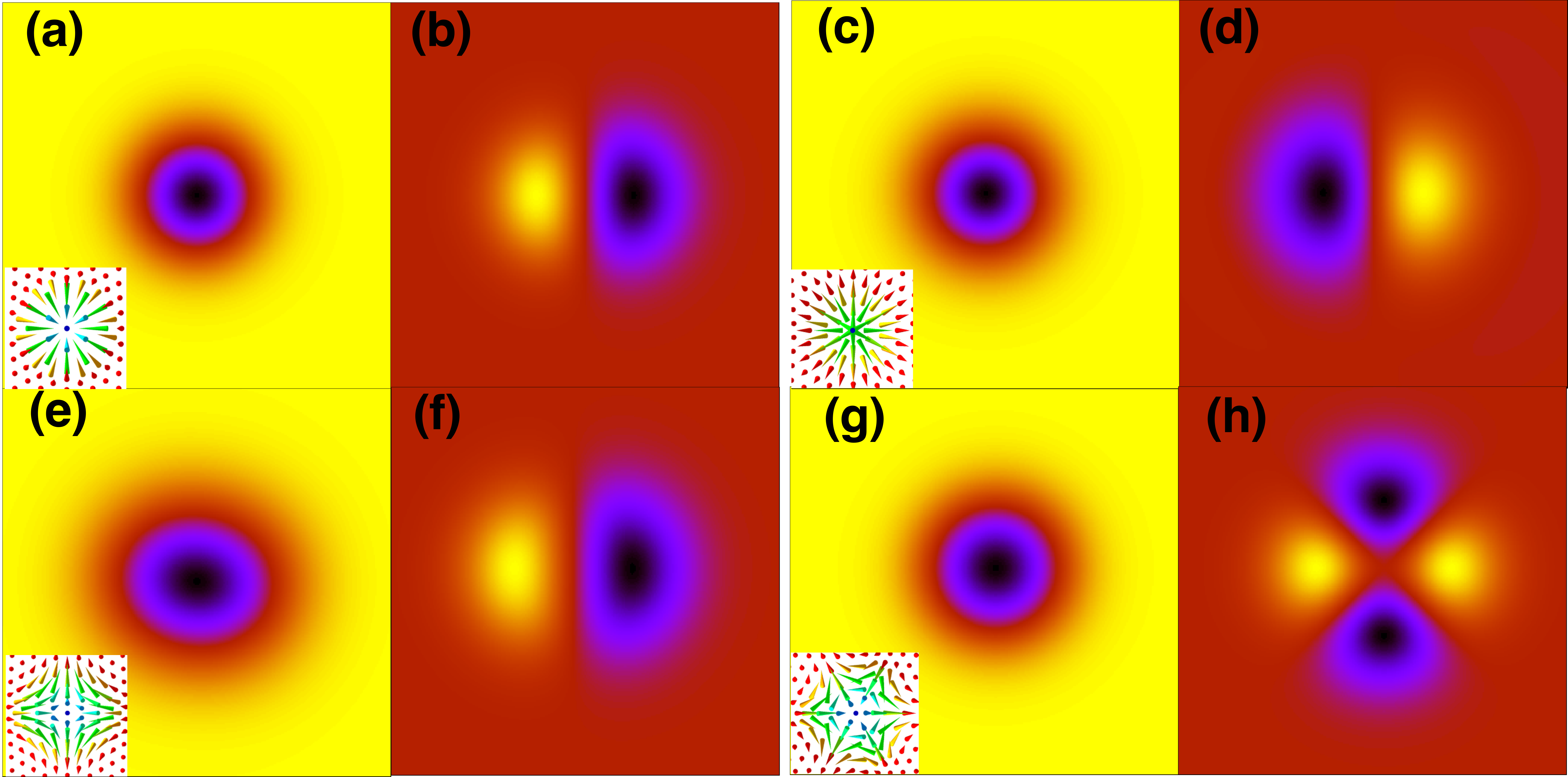}
\caption{Simulated SP-STM images for an out-of-plane magnetized tip (left column) and for an in-plane magnetized tip along the 
horizontal axis (right column). (a) and (b) right-handed skyrmion, (c) and (d) left-handed skyrmion, (e) and (f) antiskyrmion, (g) and (h) 
higher-order antiskyrmion. For all spin structures the image area as well as the color scale of the contrast are the same. The insets show the spin structures.}
\label{fig:static-STM}
\end{figure}

The simulated STM images of the isolated spin structures are shown in Fig.~\ref{fig:static-STM}.	
An out-of-plane magnetized tip (left column of Fig.~\ref{fig:static-STM}) is sensitive to the perpendicular component of the local magnetization direction. For the FM background, the tunneling current is high which creates the yellow contrast. The current decreases when the tip scans across the skyrmion due to the antiparallel spin alignment 
and the contrast darkens. All spin structures appear as spherical entities in this imaging mode.
When the tip magnetization changes to in-plane, the images of the right-handed skyrmion (Fig.\ref{fig:static-STM}(b)), the left-handed skyrmion (Fig.~\ref{fig:static-STM}(d)) 
and the antiskyrmion (Fig.~\ref{fig:static-STM}(f)) are still very similar.
The simulated SP-STM images of a skyrmion and an antiskyrmion could only differ by a rotation as seen Fig.~\ref{fig:relax}(b). The only spin structure that can be easily distinguished is the higher-order skyrmion due to the multiple nodes of the contrast 
(cf. Fig.~\ref{fig:static-STM}(h)). Note, that the simulated SP-STM images of the right-handed skyrmion for both magnetization directions
 are in good agreement with the experiments of Romming {\it et al.} \cite{Romming-2013aa,Romming2015}.

\begin{figure}[thp]
\centering
\includegraphics[width=\columnwidth]{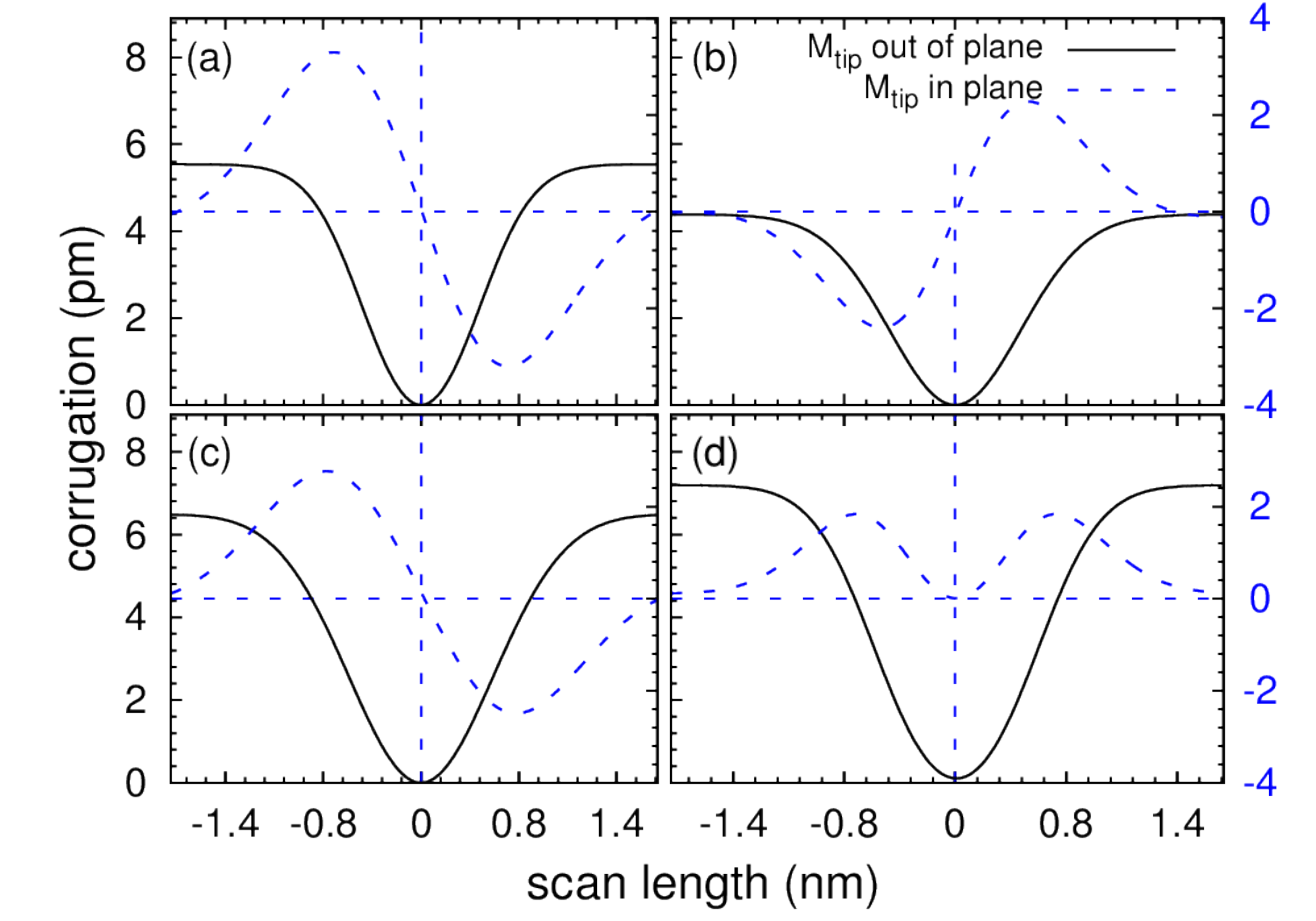}
\caption{Line profiles of the simulated SP-STM images for the topologically different spin structures of Fig.~\ref{fig:static-STM}. Solid black lines correspond to an 
out-of-plane magnetized tip and dashed blue lines correspond to an in-plane magnetized tip. Line profile of (a) the right-handed skyrmion, (b) the left-handed skyrmion, (c) the antiskyrmion and (d) the higher-order antiskyrmion.}
\label{fig:profile}
\end{figure}

The contrast of the SP-STM images can be quantified by the corrugation amplitude, i.e.~the maximal change of tip height as it is scanned 
across the spin structure, which is obtained by analyzing scan lines. 
From the simulated images one expects that such line profiles are very similar which is indeed confirmed in Fig~\ref{fig:profile}.
For an out-of-plane magnetized tip, which leads to qualitatively very similar contrasts in the SP-STM images (cf.~Fig~\ref{fig:static-STM}), 
we obtain line profiles which are also in quantitative agreement with experimental data \citep{Romming2015}.
These line profiles are characterized by a smooth change from a high tip-sample separation
in the FM background (when the magnetization of tip and the local magnetization of the sample are aligned) to a low separation at the center of the spin structures. Very similar corrugation amplitudes are found in our simulations for all spin structures, with corrugation values of 5.6, 4.4, 6.4, and 7.1 pm for the right-handed skyrmion, the left-handed skyrmion, the 
antiskyrmion and the higher-order antiskyrmion, respectively. These 
corrugation amplitudes are on the same order as found in the experiments \cite{Romming2015}.

When the tip magnetization is switched to in-plane, the line profiles show the same behavior for the right-handed skyrmion, the 
left-handed skyrmion and the antiskyrmion (Fig.~\ref{fig:profile}(a-c)). Since the 
contrast of the SP-STM images of the skyrmion and antiskyrmion are
only rotated with respect to each other and the corrugation amplitudes are very similar, they can only be distinguished in experiments when 
both spin structures are present simultaneously. 
On the other hand, the higher-order skyrmion (Fig.~\ref{fig:profile}(d)) can be easily discriminated due to the presence of multiple nodes of 
the magnetization density also seen in Fig.~\ref{fig:static-STM}(h).

\begin{figure}[thp]
\centering
\includegraphics[width=\columnwidth]{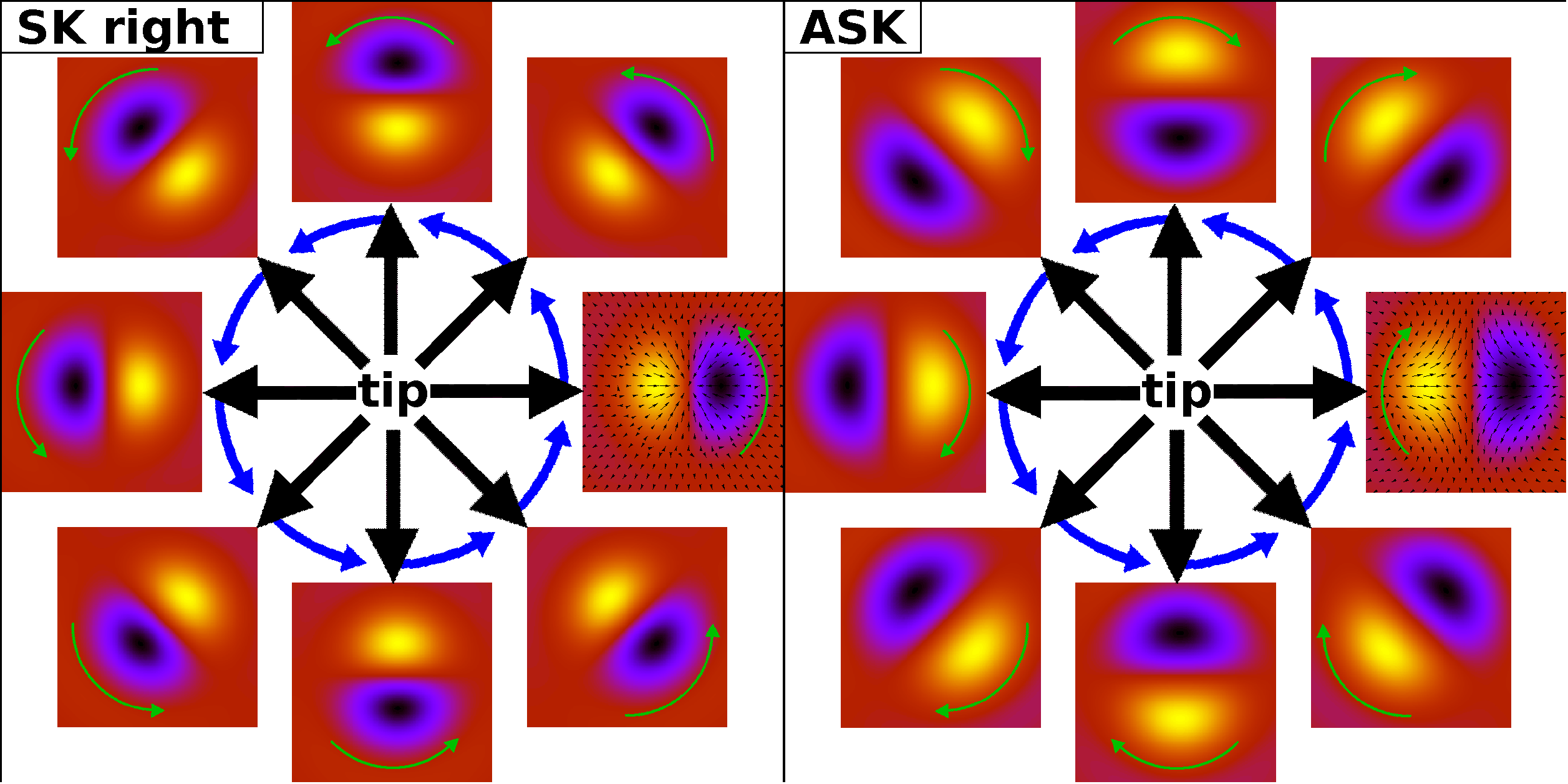}
\caption{Simulated SP-STM experiments in which the tip magnetization is rotated in the film plane.
For a right-handed skyrmion (left panel), the butterfly contrast rotates clockwise as the magnetization of the tip, while for an antiskyrmion (right panel), the 
butterfly rotates counterclockwise opposite to the tip magnetization.}
\label{fig:SPSTM-rotate}
\end{figure}

In order to distinguish between the skyrmion and the antiskyrmion spin structures, we propose an experiment based on a 3D vector field available in STM experiments \citep{Meckler:2009aa}. Such a field enables a rotation of the tip magnetization both within the surface plane as well as from in-plane to out-of-plane. 
Although all in-plane magnetized tips result in the same contrast i.e.~a butterfly with a bright and a dark lobe (as shown in Fig.~\ref{fig:static-STM}), the behavior of this contrast when the tip changes its direction is different as shown in Fig.~\ref{fig:SPSTM-rotate}. For a right-handed skyrmion, the lobes will rotate in phase with the tip
magnetization direction (thick black arrows). In the case of an antiskyrmion, a clockwise rotation of the in-plane component of the tip will induce a counterclockwise rotation of the lobes. Therefore, in-plane rotation of the tip magnetization allows an unambiguous distinction between the right-handed skyrmion and the antiskyrmion. 
On the other hand, in order to distinguish a left- and right-handed skyrmion the tip magnetization must be rotated from the in-plane to the out-of-plane direction.

\section{Conclusion}

In conclusion, we have demonstrated that it is non-trivial to distinguish via SP-STM between metastable spin structures at surfaces that differ by their chirality and/or 
topological charge. Skyrmions and antiskyrmions exhibit a spherical shape in SP-STM using tips with an out-of-plane magnetization. For in-plane magnetized tips we obtain
a characteristic butterfly pattern that is aligned along the tip magnetization for skyrmions, while the alignment depends on the antiskyrmion orientation with respect
to the tip magnetization. If the tip magnetization is rotated within the surface plane the butterfly contrast rotates in phase with the tip direction for skyrmions 
and in the opposite direction for antiskyrmions allowing to unambiguously distinguish between them. The demonstration of stabilizing localized spin structures with different chirality and topological charge in the same system e.g. Pd/Fe/Ir(111), opens new possibilities for spintronics \cite{Koshibae2016}.
 
\section*{Acknowledgments}
B. Dup\'e and S. Heinze thank the Deutsche Forschungsgemeinschaft (DFG) for financial support via the project DU 1489/2-1 
and gratefully acknowledge computing time at the HLRN supercomputer. This project has received funding from the European Unions Horizon 2020 research and innovation pro- gramme under grant agreement No 665095.


\begin{thebibliography}{28}
\bibitem [{coe()}]{coeff}%
  \BibitemOpen
  \href@noop {} {\bibinfo  {journal} {\normalfont We have obtained the
  coefficients of the spin Hamiltonian from first-principles calculations for
  Pd(fcc)/Fe/Ir(111) [21]. The exchange constants $J_n$ between the
  $n$thnearest-neighbors in the Fe layer $J_1 \cdots J_{10}$ are given, in meV,
  by $14.73$, $-1.95$, $-2.88$, $0.32$, $0.69$, $0.01$, $0.01$, $0.13$,
  $-0.14$, $-0.28$, respectively, the anisotropy constant is $K=-0.7$ meV,
  which corresponds to an out-of-plane easy axis and the DMI constant is
  $D=1.0$ meV.}\ }\BibitemShut {NoStop}%
\end{thebibliography}
\end{document}